\documentclass[a4paper,11pt]{article}

\usepackage[utf8]{inputenc}
\usepackage{amsmath}
\usepackage{tikz}
\usepackage{verbatim}
\usepackage{pgffor}
\usepackage{listings}
\usepackage{placeins}
\usepackage{float}
\usepackage{cite}
\usepackage{algorithm,algorithmic}
\usepackage{color}
\usepackage{jheppub}
\usepackage{tcolorbox}
\usepackage[thinlines]{easytable}
\usepackage{cite}
\usetikzlibrary{decorations.pathmorphing}
\usetikzlibrary{decorations.markings}
\usetikzlibrary{shapes.geometric}
\usepackage{cancel}
\usepackage{subcaption}
\usepackage{url}

\tikzset{snake it/.style={decorate, decoration=snake}}
\tikzset{wavy/.style={decorate, decoration={complete sines,amplitude=1.5pt, segment length=4pt}}}
\usetikzlibrary{matrix}
\usetikzlibrary{arrows.meta}
\usetikzlibrary{shapes.geometric, arrows}
\tikzstyle{startstop} = [rectangle, rounded corners, minimum width=3cm, minimum height=1cm,text centered, draw=black, fill=red!30]
\tikzstyle{io} = [trapezium, trapezium left angle=70, trapezium right angle=110, minimum width=3cm, minimum height=1cm, text centered, draw=black, fill=blue!30]
\tikzstyle{process} = [rectangle, minimum width=3cm, minimum height=1cm, text centered, draw=black, fill=orange!30]
\tikzstyle{decision} = [diamond, minimum width=3cm, minimum height=1cm, text centered, draw=black, fill=green!30]
\tikzstyle{arrow} = [thick,->,>=stealth]
\tikzstyle arrowstyle=[scale=1]
\tikzstyle directed=[postaction={decorate,decoration={markings,
    mark=at position .65 with {\arrow[arrowstyle]{stealth}}}}]

\makeatletter
\def\maketag@@@#1{\hbox{\m@th\normalfont\normalsize#1}}
\makeatother

\def\eps{\epsilon}

\def\tr{\text{Tr}}
\def\PT{\text{PT}}

\newcommand{\vev}[1]{\ensuremath{\langle #1 \rangle }}

\newcommand{\ab}[2]{\ensuremath{\langle #1 \, #2 \rangle}}
\newcommand{\sqb}[2]{\ensuremath{\lbrack #1 \, #2 \rbrack}}
\newcommand{\spn}[1]{\lambda_{#1}}
\newcommand{\asp}[1]{\tilde{\lambda}_{#1}}

\title{Conformal Invariance of the One-Loop All-Plus Helicity Scattering Amplitudes}
\author[a]{Johannes Henn}
\author[a,b]{Bl\'aith\'in Power}
\author[a]{Simone Zoia}

\affiliation[a]{Max-Planck-Institut f\"{u}r Physik, Werner-Heisenberg-Institut, D-80805 M\"{u}nchen, Germany}
\affiliation[b]{Ludwig-Maximilians-Universit\"{a}t, Theresienstra{\ss}e 37, 80333 M\"{u}nchen, Germany}

\emailAdd{henn@mpp.mpg.de}
\emailAdd{powerbl@mpp.mpg.de}
\emailAdd{zoia@mpp.mpg.de}

\preprint{MPP-2019-238}

\bigskip

\abstract{The massless QCD Lagrangian is conformally invariant and, as a consequence, so are the tree-level scattering amplitudes.
However, the implications of this powerful symmetry at loop level are only beginning to be explored systematically. Even for finite loop amplitudes, the way conformal symmetry manifests itself may be subtle, e.g. in the form of anomalous conformal Ward identities. As they are finite and rational, the one-loop all-plus and single-minus amplitudes are a natural first step towards understanding the conformal properties of Yang-Mills theory at loop level.
Remarkably, we find that the one-loop all-plus amplitudes are conformally invariant, whereas the single-minus are not. 
Moreover, we present a formula for the one-loop all-plus amplitudes where the symmetry is manifest term by term.
Surprisingly, each term transforms covariantly under directional dual conformal variations.
We prove the formula directly using recursive techniques, and check that it has the correct physical factorisations.}

\begin{document}
\maketitle

\newpage

\section{Introduction}
\label{sec:Introduction}

Particle collisions take place at the Large Hadron Collider  at extremely high energies. As a consequence, the masses of the scattered particles can often be neglected, in which case (the Lagrangian of) the Standard Model becomes conformally invariant at classical level. 

The implications of conformal symmetry have been extensively explored in position space, where it found many successful applications in the calculation of anomalous dimensions and correlation functions, or in the search for consistency conditions constraining the space of conformal field theories through the operator product expansion. This is a very active field of research. However, much less is known about the consequences of conformal symmetry for scattering amplitudes, which live in on-shell momentum space.

Scattering amplitudes play a crucial role in quantum field theory. They connect theory and experiment, as they are the most basic ingredient in the computation of cross sections. Moreover, being the gauge-invariant building blocks for cross sections, their study often gives precious insights into the structure of the underlying theory. For instance, the dual (super)conformal symmetry of planar $\mathcal{N}=4$ super Yang-Mills is nowhere to be seen in the Lagrangian, and emerges only at the level of the scattering amplitudes~\cite{Drummond:2008vq}.

However, studying the consequences of conformal symmetry for scattering amplitudes is challenging for several reasons. 
Primarily, scattering amplitudes in quantum field theory are typically divergent at quantum level. The necessity of regulating infrared and ultraviolet divergences forces one to introduce a mass scale, which obscures the conformal symmetry of the Lagrangian. This is a major issue, but there is hope. Infrared and ultraviolet divergences factorise separately in a well understood way~\cite{Collins:1989gx}. It is possible to subtract them, and define a finite ``hard part" of the amplitude, in which the regulator can be removed. It is therefore not unreasonable to speculate that it may be possible to find a particular subtraction and renormalisation scheme in which this hard function exhibits, in some way, the underlying conformal properties.  Another approach, following~\cite{Braun:2003rp,Braun:2017cih} and references therein, is to study the theory at a conformal fixed point.

One might bypass the fundamental problem of divergences by studying finite loop amplitudes, only to find that the idea of conformal symmetry being broken at loop level only by the dimensionful regulators is actually na\"ive.
Indeed, even for finite loop integrals that are exactly conformally invariant off-shell, the way conformal symmetry is implemented on-shell can be very subtle. In Refs.~\cite{Bargheer:2009qu,Korchemsky:2009hm,Beisert:2010gn,Chicherin:2017bxc,Chicherin:2018ubl,Chicherin:2018rpz}, it was shown that on-shell effects can lead to conformal Ward identities with a non-zero anomaly term.

On top of all this, there is also a practical complication. The generator of special conformal transformations, which is a first-order differential operator in coordinate space, becomes second-order in momentum space. This means that, even to answer the most basic questions --~for instance, what is the most general conformal invariant for a given particle configuration?~-- requires the solution of a system of second-order partial differential equations. As the number of variables increases, this soon becomes a formidable task.
Hence the importance of finding, by any means, explicit examples of conformal invariants, to understand and reproduce the mechanism underlying their symmetry.

Despite all the issues, the study of the implications of conformal symmetry for scattering amplitudes is not only of deep theoretical interest, but is also rewarding from the practical point of view. Although a systematic investigation has been started only recently, we can already quote several applications. The anomalous conformal Ward identities of Refs.~\cite{Chicherin:2017bxc,Chicherin:2018ubl,Chicherin:2018rpz,Zoia:2018sin}, for instance, have been used to compute non-trivial loop integrals, including complicated two-loop five-particle examples. Conformal symmetry can be used also in relation to the rational functions appearing in a scattering amplitude. The on-shell diagrams~\cite{ArkaniHamed:2012nw} built by assembling conformally invariant vertices are conformally invariant in their turn. Knowing the conformal objects for a given particle configuration therefore gives a handle on the leading singularities~\cite{Cachazo:2008vp} of the corresponding scattering amplitudes, which are related to rational factors appearing in them. 

For example, the rational factors in the polylogarithmic part of the two-loop all-plus amplitudes have been computed for arbitrary number of gluons in the planar limit using four-dimensional unitarity~\cite{Dunbar:2016cxp}. In this approach, the one-loop all-plus amplitudes have no cuts and may be regarded as vertices. The rational factors are then given by on-shell diagrams which involve tree-level amplitudes and the one-loop all-plus amplitudes as vertices. The conformal invariance of the one-loop amplitudes, which we prove in this paper, straightforwardly implies that of these rational factors. It is natural to expect the same to hold for the non-planar contributions. Indeed, in the recently computed full two-loop five-gluon all-plus amplitude~\cite{Badger:2019djh,Dunbar:2019fcq}, the polylogarithmic part contains the same conformally invariant rational factors which appear in our formula for the one-loop amplitude.

The on-shell effects breaking (super)conformal symmetry for finite (super)integrals have been studied in scalar and Yukawa theories, and matter supermultiplet~\cite{Chicherin:2017bxc,Chicherin:2018ubl,Chicherin:2018rpz,Zoia:2018sin}. 
We aim to make the first step towards understanding the conformal properties of Yang-Mills theory at loop level.
In any supersymmetric theory, the gluon scattering amplitudes with no or one negative helicity gluon vanish. Given this, they are the natural first objects to study in a general Yang-Mills theory. Since they vanish at tree-level, they are finite and rational at one loop. The one-loop all-plus amplitudes meet the na\"ive expectation of being conformally invariant. The single-minus amplitudes do not\footnote{We have checked that the one-loop single-minus amplitudes are not conformally invariant for \mbox{$n=4,5$}~\cite{Bern:1991aq,Bern:1993mq}.}. 

In this work, we prove that the one-loop all-plus helicity amplitudes are conformally invariant for any number of gluons. We do this by deriving a new formula in which the symmetry is manifest term by term. 
We present and discuss the new expression in Section~\ref{sec:PresentationResult}. Its manifest conformal symmetry is proven in Section~\ref{sec:ConformalSymmetry}. Remarkably, the new formula has non-trivial properties under dual conformal symmetry as well, which we highlight in Section~\ref{sec:DualConformalSymmetry}. Section~\ref{sec:FactorizationSpurious} is then devoted to the analysis of the analytic properties: we show that the new formula has the correct leading behaviour in the soft and collinear limits, and that it is free of spurious poles.
In Section~\ref{sec:BCFW} we prove our formula by combining two complementary BCFW recursions.
We draw our conclusions and discuss the future research directions in Section~\ref{sec:Conclusions}.

\section{Manifestly conformal result for the one-loop all-plus amplitude in QCD}
\label{sec:PresentationResult}

The $n$-gluon all-plus amplitudes vanish at tree level, and are given by finite and rational functions at one loop~\cite{Bern:1993sx, Mahlon:1993si, Bern:1993qk}. 
In this section we present a new, manifestly conformally invariant formula for the one-loop all-plus amplitudes. 

We start by defining our conventions. We expand the $n$-gluon all-plus helicity scattering amplitude in powers of the coupling constant $g$ as
\begin{align}
\label{eq:PerturbativeExpansion}
\mathcal{A}_{n}(\{p_i, a_i \}) = \frac{g^{n}}{16 \pi^2} A_{n}^{(1)}(\{p_i, a_i \}) \, \delta^{(4)}(p_1 + \ldots + p_n)  + \mathcal{O}(g^{n+2}) \, ,
\end{align}
where $p_i$ and $a_i$ are the momentum and adjoint colour index of the $i$-th gluon.
In order to make the colour dependence explicit, we adopt the following decomposition into colour-ordered partial amplitudes~\cite{Bern:1990ux},
\begin{align}
\label{eq:ColorDecomposition}
A_n^{(1)}&(\{p_i, a_i \}) \, = \sum_{\sigma \in S_n/Z_n} N_c \, \tr\left(T^{a_{\sigma(1)}}\ldots T^{a_{\sigma(n)}} \right) A^{(1,1)}_{n}\left(\sigma(1^+),\ldots, \sigma(n^+)\right) + \nonumber \\
& + \sum_{c=2}^{\left \lfloor{n/2}\right \rfloor  + 1} \sum_{\sigma\in S_n/S_{n;c}}\tr\left(T^{a_{\sigma(1)}}\ldots T^{a_{\sigma(c-1)}} \right) \tr\left(T^{a_{\sigma(c)}}\ldots T^{a_{\sigma(n)}} \right) A^{(1,0)}_{n}\left(\sigma(1^+),\ldots, \sigma(n^+)\right) \, ,
\end{align}
where $S_n$ is the set of all permutations of $n$ objects, $Z_n$ is the subset of cyclic permutations leaving the single trace invariant, and $S_{n;c}$ is the subset which leaves the corresponding double trace invariant.
The $T^{a_i}$ are the generators of the $SU(N_c)$ gauge group in the fundamental representation, normalised such that $\tr\left(T^a T^b \right) = \delta^{ab}$. 

The double-trace components $A^{(1,0)}_n$  in Eq.~\eqref{eq:ColorDecomposition} are determined by the single-trace ones $A^{(1,1)}_n$  through $U(1)$ decoupling relations~\cite{Bern:1990ux}. We will therefore focus on the latter. 

The single-trace components $A^{(1,1)}_n$ are colour ordered, namely they receive contributions only from planar graphs where the cyclic ordering of the external legs is fixed to match that of the generators in the corresponding trace. Because of this, they are called planar (partial) amplitudes, and their singularities can occur only in channels of adjacent momenta. Note that the complete amplitude is invariant under permutations of the external legs. From this, thanks to the cyclic symmetry of the traces, the planar partial amplitudes $A_{n}^{(1,1)}$ inherit symmetry under cyclic permutations of the external legs.

In QCD with $n_f$ quarks and gauge group $SU(N_c)$, the planar one-loop all-plus amplitudes in four dimensions take a remarkably compact form~\cite{Bern:1993sx, Mahlon:1993si, Bern:1993qk},
\begin{align}\label{allplus_all_n}
A^{(1,1)}_{n}(1^+,\ldots, n^+) = \frac{M}{\PT(1,2, \ldots,n)} \sum_{1\le i_1<i_2<i_3<i_4 \le n}  \langle i_{1} i_{2} \rangle [ i_{2} i_{3}] \langle i_{3} i_{4} \rangle [ i_{4} i_{1}]    \,,
\end{align}
where $M$ is an overall constant factor,
\begin{align} M = - \frac{i}{3} \left( 1- \frac{n_f}{N_c} \right)\, , \end{align}
and we defined the Parke-Taylor denominator as
\begin{align}
\PT(i_1, i_2,\ldots, i_n) = \langle i_1 i_2 \rangle \langle i_2 i_3 \rangle \ldots \langle i_{n-1} i_n \rangle \langle i_n i_1 \rangle \,.
\end{align}
Equation~\eqref{allplus_all_n} was conjectured in Refs.~\cite{Bern:1993sx, Bern:1993qk} based on the behaviour as external momenta become collinear~\cite{PhysRevLett.56.2459,BERENDS1988759,MANGANO1988673}. The conjecture was then proven in Ref.~\cite{Mahlon:1993si} by off-shell recursion relations~\cite{BERENDS1988759}.

In this paper we derive a different expression,\footnote{This formula appeared previously in Refs.~\cite{Mafra:2012kh,He:2015wgf}, in the context of string theory amplitudes. We thank the authors of~\cite{He:2015wgf} for bringing these references to our attention.}
\begin{align}\label{eq:our_formula}
A_{n}^{(1,1)}\left(1^+, \ldots, n^+ \right)  = M \,  \sum_{k=3}^{n-1}  \sum_{m=k+1}^{n}  C_{kmn}\,,
\end{align}
where
\begin{align}
\label{eq:Ckmn}
C_{kmn} = - \frac{ \langle 1  | x_{1k} x_{km} | 1 \rangle^2}{\PT(1, 2,  \ldots, k-1 ) \, \PT(1, k, k+1, \ldots, m-1 ) \, \PT(1, m, m+1, \ldots , n ) } \,,
\end{align}
with the dual coordinates defined as
\begin{align}
\label{eq:xab}
x_{ab} = x_a - x_b = p_{a} + p_{a+1} + \ldots + p_{b-1} \, ,
\end{align}
and $\langle 1  | x_{1k} x_{km} | 1 \rangle = \lambda_1^{\alpha} (x_{1k})_{\alpha \dot{\alpha}} (x_{km})^{\dot{\alpha} \beta} \lambda_{1 \, \beta}$.

As we will show in Section~\ref{sec:ConformalSymmetry}, the summands $C_{kmn}$ in our formula~\eqref{eq:our_formula} are individually conformally invariant. This implies in a very neat way that the one-loop all-plus amplitude is conformally invariant for any number of external gluons. This is not obvious at all in the original formula~\eqref{allplus_all_n}, where the summands are not separately conformally invariant.

Moreover, it is intriguing to note that our formula~\eqref{eq:our_formula} is reminiscent of similar formulas for NMHV super-amplitudes in $\mathcal{N}=4$ super Yang-Mills~\cite{Drummond:2008vq}. In particular, the $C_{kmn}$ show some hints of dual conformal symmetry, which can be highlighted by rewriting them in another way,
\begin{align}
\label{eq:CkmnAlternative}
C_{kmn}=&- \frac{1}{\PT(1,\ldots, n)} \frac{ \langle 1 | x_{1k} x_{km} | 1 \rangle^2 \langle k-1 \, k\rangle \langle m-1\, m \rangle}{\langle 1\, k-1\rangle \langle 1 k \rangle \langle 1\, m-1\rangle \langle 1 m \rangle}  \, .
\end{align}
The dual conformal properties of the summands $C_{kmn}$ will be discussed in Section~\ref{sec:DualConformalSymmetry}.

The vanishing of the all-plus and single-minus tree-level amplitudes implies not only that the one-loop all-plus amplitude is finite and rational, but also that it does not contain multi-particle poles. Only the singularities where two (colour-)adjacent momenta become collinear are allowed. While this is clear in Eq.~\eqref{allplus_all_n}, we will show in Section~\ref{sec:FactorizationSpurious} that the same holds for Eq.~\eqref{eq:our_formula} as well. 

Finally, note that the cyclic symmetry is not manifest in Eq.~\eqref{eq:our_formula}. The gluon with momentum $p_1$, for instance, appears to have a somewhat special role in the summands~\eqref{eq:Ckmn}, which is tightly linked to the dual conformal symmetry properties discussed in Section~\ref{sec:DualConformalSymmetry}. This is just an artefact of the specific representation we chose, and different choices are of course possible.

\section{Proof of manifest conformal symmetry}
\label{sec:ConformalSymmetry}

The tree-level gluon scattering amplitudes are conformally invariant. This means that, on top of enjoying Poincaré symmetry, they are invariant under dilatations and special conformal transformations. While the invariance under dilatations is guaranteed by the absence of dimensionful parameters, that under special conformal transformations is more interesting. It means that the tree-level gluon amplitudes are annihilated by the second-order generator~\cite{Witten:2003nn}\footnote{Note that we are interested in generic momentum configurations, and we therefore neglect contact terms arising from differentiation~\cite{Witten:2003nn,Cachazo:2004by,Bargheer:2009qu,Korchemsky:2009hm,Beisert:2010gn}. 
Furthermore, the amplitudes always contain an overall momentum conservation $\delta$ function. Since the amplitudes are Lorentz and dilatation invariant, the generator $k_{\alpha \dot{\alpha}}$ commutes with the $\delta$ function~\cite{Witten:2003nn,Chicherin:2017bxc}.}
\begin{align}\label{conformal_generator}
k_{\alpha \dot{\alpha}} = \sum_{i=1}^{n} \frac{\partial^2}{\partial \lambda_{i}^{\alpha} \partial \tilde{\lambda}_{i}^{\dot{\alpha}}} \, .
\end{align}

The all-plus amplitudes vanish at tree-level, and are therefore finite at one loop. The absence of divergences, and thus of dimensionful scales to treat them, suggests that conformal symmetry might survive at one loop in this special case. Recent studies~\cite{Chicherin:2017bxc,Chicherin:2018ubl,Chicherin:2018rpz,Zoia:2018sin} have however warned us against very subtle conformal symmetry breaking mechanisms even for finite loop integrals. Indeed, the single-minus amplitudes, which are finite at one-loop just like the all-plus ones, are not conformally invariant.

In this section we show that the all-plus amplitudes are conformally invariant at one loop. 
We begin by spelling out some low-multiplicity cases. Then we will prove analytically that our formula~\eqref{eq:our_formula} is conformally invariant term by term for any number of external gluons.

Let us begin with the simplest case: the four-gluon amplitude. From Eq.~\eqref{allplus_all_n}, for $n=4$, we have 
\begin{align}\label{eq:A4literature}
A_{4}^{(1,1)}\left(1^+, 2^+, 3^+, 4^+\right) = M \, \frac{[23][41]}{\vev{23} \vev{41} } \,.
\end{align}
Using momentum conservation and Schouten identities, this can be rewritten in many equivalent ways. In particular, 
\begin{align}\label{A4explicit}
A_{4}^{(1,1)}\left(1^+, 2^+, 3^+, 4^+\right) = M \, \frac{[23]^2}{\langle 14 \rangle^2} \,.
\end{align}
This expression, although exhibiting spurious double poles, has the advantage of displaying in a spectacularly manifest way its conformal invariance, 
\begin{align}
k_{\alpha \dot\alpha} \, A_{4}^{(1,1)}\left(1^+,  2^+, 3^+, 4^+\right) = 0\,.
\end{align}
In fact, no calculation is needed to show it. If we recall the form of the conformal generator~\eqref{conformal_generator}, it contains a second-order mixed partial derivative with respect to $\spn{i}$ and $\asp{i}$ for all particles $i = 1,\ldots,4$. The expression~\eqref{A4explicit} contains only $\lambda$ or $\tilde{\lambda}$ spinors for each particle, and it is therefore trivially annihilated.

Let us now look at the case $n=5$. Five-particle amplitudes have a special feature: they depend on a pseudo-scalar invariant, $\eps_5 = 4  i \epsilon_{\mu \nu\rho\sigma} p_1^{\mu} p_2^{\nu} p_3^{\rho} p_4^{\sigma}$, which vanishes in all collinear limits. While the constraints on the spurious poles and the collinear limits fix entirely the form of the one-loop all-plus amplitudes for $n>5$~\cite{Bern:1993sx, Bern:1993qk}, the dependence on $\epsilon_5$ at $n=5$ remains elusive. Quite remarkably, conformal symmetry fixes it as well, once the rest of the amplitude is given. 
The five-gluon amplitude is usually presented in the following way~\cite{Bern:1993mq},
\begin{align}
\label{eq:A5original}
A_{5}^{(1,1)}\left(1^+,\ldots, 5^+\right) = - \frac{M}{2} \frac{s_{12} s_{23} + s_{23} s_{34} + s_{34} s_{45} + s_{45} s_{51}+ s_{51} s_{12} + \eps_5}{\PT(1,\ldots,5)} \, .
\end{align}
There is no linear combination of the terms appearing in Eq.~\eqref{eq:A5original} that is conformally invariant, other than the one which constitutes the amplitude.

As noticed in Ref.~\cite{Badger:2019djh}, the five-gluon all-plus amplitude can be rewritten in a manifestly conformally invariant way, 
\begin{align}\label{A5explicit}
A_{5}^{(1,1)}\left(1^+,\ldots, 5^+\right) = - M \, \left( \frac{[45]^2}{\langle 12 \rangle \langle 23 \rangle \langle 31 \rangle } + \frac{[23]^2}{\langle 45 \rangle \langle 51 \rangle \langle 14 \rangle } + \frac{[52]^2}{\langle 41 \rangle \langle 13 \rangle \langle 34 \rangle }  \right) \,.
\end{align}
Each term in this expression is trivially annihilated by the conformal generator~\eqref{conformal_generator}, in the same way the mechanism worked at $n=4$.

Comparing the collinear properties of the original~\eqref{eq:A5original} and of the new~\eqref{A5explicit} formula reveals a tension between manifest conformal symmetry and manifest analytic properties. While Eq.~\eqref{eq:A5original} obviously contains poles only where two colour-adjacent momenta become collinear, Eq.~\eqref{A5explicit} exhibits spurious poles. What is more, although Eq.~\eqref{A5explicit} can be rewritten in several manifestly conformally invariant ways, none of them is free of spurious poles. In fact, there are ${5\choose 2} = 10$ conformal objects of the form $[ij]^2/(\vev{ab}\vev{bc}\vev{ca})$ with $i,j,a,b,c$ all different, and each one of them contains necessarily at least one spurious pole. Very interestingly, the residues at the spurious poles vanish in a beautiful way thanks to cancellations between different conformal invariants. This can be shown for arbitrary multiplicity, as we discuss in Section~\ref{sec:FactorizationSpurious}.

Some readers might recognise that the appearance of spurious poles is a typical feature of BCFW representations. Indeed, we will show in Section~\ref{sec:BCFW} that our formula~\eqref{eq:our_formula} is constructed from a suitably chosen BCFW recursion. 

So far, the mechanism allowing for conformal symmetry has been trivial. More interesting conformally invariant objects appear starting from $n=6$. In this case, our formula~\eqref{eq:our_formula} gives
\begin{align}\label{eq:A6explicit}
A_{6}^{(1,1)}\left(1^+,\ldots, 6^+\right) = M \biggl( - \frac{[23]^2}{\vev{14}\vev{45}\vev{56}\vev{61}} - \frac{[26]^2}{\vev{13}\vev{34}\vev{45}\vev{51}} - \frac{[56]^2}{\vev{12} \vev{23}\vev{34} \vev{41}} + \nonumber \\
+ \frac{\langle 1| p_3+p_4 | 2]^2}{\vev{13} \vev{34} \vev{41} \vev{15} \vev{56} \vev{61}} + \frac{\langle 1| p_5+p_6 | 4]^2}{\vev{12} \vev{23} \vev{31} \vev{15} \vev{56} \vev{61}} + \frac{\langle 1| p_2+p_3 | 6]^2}{\vev{12} \vev{23} \vev{31} \vev{14} \vev{45} \vev{51}} \biggr) \, .
\end{align}
In the first line, we recognise the generalisation of the terms already seen in the $n=4,5$ cases, together with new objects in the second line. The conformal invariance of the latter is less obvious, and was proven in Ref.~\cite{EdwardWangBSc}.

We now proceed to prove analytically that each summand $C_{kmn}$ in Eq.~\eqref{eq:our_formula} is individually conformally invariant for generic multiplicity $n$.
In order to do this, it is convenient to write it as
\begin{align}
\label{eq:rewriting}
C_{kmn} = \frac{ C_{kmn \, \dot{\beta} \dot{\sigma}}^{(1)} \, C_{kmn}^{(2) \, \dot{\beta} \dot{\sigma}}  }{\PT(1,m,\ldots,n)} \, ,
\end{align}
with
\begin{align}
C_{kmn \, \dot{\beta} \dot{\sigma}}^{(1)} = \frac{(\langle 1|x_{1k})_{\dot{\beta}} (\langle 1|x_{1k})_{\dot{\sigma}}}{\PT(1,\ldots, k-1)} \, , \quad \qquad C_{kmn}^{(2) \, \dot{\beta} \dot{\sigma}} = \frac{(x_{km}|1\rangle)^{\dot{\beta}} (x_{km}|1\rangle)^{\dot{\sigma}}}{\PT(1,k,\ldots, m-1)} \, .
\end{align}
This rewriting in fact exposes a separation in the dependence on the particles. Particles $\{1,m,m+1,\ldots,n\}$ appear only in a holomorphic way. Their contribution to the generator $k_{\alpha \dot{\alpha}}$~\eqref{conformal_generator} therefore vanishes trivially. 
The remaining particles are then divided into two sets, $\{2, \ldots,k-1\}$ and $\{k,\ldots,m-1\}$. Since $C_{kmn \, \dot{\beta} \dot{\sigma}}^{(1)}$ depends only on the former set, and $C_{kmn}^{(2) \,  \dot{\beta} \dot{\sigma}}$ only on the latter, the action of the generator of $k_{\alpha \dot{\alpha}}$~\eqref{conformal_generator} splits. Therefore, to show the invariance of Eq.~\eqref{eq:rewriting} it is sufficient to prove that $k_{\alpha \dot{\alpha}} C^{(1) \dot{\beta}\dot{\sigma}}_{kmn} = 0$ and $k_{\alpha \dot{\alpha}} C^{(2) \dot{\beta}\dot{\sigma}}_{kmn} = 0$.

We show here in some detail the first computation:
\begin{equation}
\begin{aligned}
\label{eq:kC1part1}
k_{\alpha \dot{\alpha}} & C_{kmn \, \dot{\beta} \dot{\sigma}}^{(1)} = - \epsilon_{\dot{\alpha} \dot{\beta}} \sum_{a=2}^{k-1} \frac{\partial}{\partial \lambda_a^{\alpha}} \biggl(  \frac{ \langle 1a\rangle \, (\langle 1|x_{1k})_{\dot{\sigma}}}{\PT(1,\ldots, k-1)}  \biggr) + \left(\dot{\beta} \leftrightarrow \dot{\sigma} \right) = \\
& = - \frac{\epsilon_{\dot{\alpha} \dot{\beta}} }{\PT(1,\ldots,k-1)} \sum_{a=2}^{k-1} \biggl\{ - \langle 1 a \rangle \tilde{\lambda}_{a \dot{\sigma}} \lambda_{1 \alpha}  -\lambda_{1  \alpha} (\langle 1|x_{1k})_{\dot{\sigma}} + \\
& \qquad + (\langle 1|x_{1k})_{\dot{\sigma}} \, \vev{1a} \left( \frac{\lambda_{a-1 , \alpha}}{\langle a-1 , a \rangle} -  \frac{\lambda_{ a+1 , \alpha}}{\langle a , a+1 \rangle} \right)\biggl|_{\text{mod} \, k-1} \biggr\} + \left(\dot{\beta} \leftrightarrow \dot{\sigma} \right) \, ,
\end{aligned}
\end{equation}
where the indices $\dot{\beta}$ and $\dot{\sigma}$ are symmetrised, and the labels of the spinors in the parentheses are defined modulus $k-1$, namely $\lambda_k^{\alpha} \bigl|_{\text{mod} \, k-1}  = \lambda_1^{\alpha} $. Next, we apply a Schouten identity and perform standard manipulations to draw out a telescoping sum,
\begin{equation}
\label{eq:kC1part2}
\begin{aligned}
k_{\alpha \dot{\alpha}} C_{kmn \, \dot{\beta} \dot{\sigma}}^{(1)} = \ & -  \frac{\epsilon_{\dot{\alpha} \dot{\beta}} \, (\langle 1| x_{1k})_{\dot{\sigma}}  + \epsilon_{\dot{\alpha} \dot{\sigma}} \, (\langle 1| x_{1k})_{\dot{\beta}} }{\PT(1,\ldots,k-1)}  \biggl\{ - \lambda_{1 \alpha}  +  \\
&  -  \sum_{a=2}^{k-1} \left(  \frac{\langle 1 a\rangle }{\langle a , a+1 \rangle} \lambda_{ a+1 , \alpha}- \frac{\langle 1 , a-1 \rangle}{\langle a-1 , a \rangle} \lambda_{a \alpha}  \right)\biggl|_{\text{mod} \, k-1} \biggr\} \, .
\end{aligned}
\end{equation}
In the second line of Eq.~\eqref{eq:kC1part2} we recognise the telescoping sum
\begin{align}
\label{eq:sum}
\sum_{a=2}^{k-1} \left( f_{a, \alpha}  -  f_{a-1 , \alpha} \right) = f_{k-1, \alpha}-f_{1, \alpha} \, ,
\end{align}
with $f_{a, \alpha} =  \frac{\langle 1 a \rangle}{\langle a , a+1 \rangle} \lambda_{a+1 , \alpha}\bigl|_{\text{mod} \, k-1} $. Since $f_{1,\alpha} = 0$ and $f_{k-1,\alpha} = - \lambda_{1\alpha}$, the sum exactly cancels out the first line of Eq.~\eqref{eq:kC1part2}, thus giving
\begin{align}
k_{\alpha \dot{\alpha}} C_{kmn \, \dot{\beta} \dot{\sigma}}^{(1)} = 0\, .
\end{align}

It is sufficient to trade the labels $\{2,\ldots, k-1\}$ for $\{k,\ldots, m-1\}$ in the chain of equalities given by Eqs.~\eqref{eq:kC1part1},~\eqref{eq:kC1part2} and~\eqref{eq:sum} to show that 
\begin{align}
k_{\alpha \dot{\alpha}} C_{kmn}^{(2) \,  \dot{\beta} \dot{\sigma}} = 0\, ,
\end{align}
which completes the proof.

Note that the conformal invariance of the planar amplitudes $A^{(1,1)}_n$ implies that of the subleading-colour ones $A^{(1,0)}_n$. Although the reverse is not true, it is still very interesting that the remarkably compact expressions for the subleading-colour components presented in Ref.~\cite{Dunbar:2019fcq} were proven to be conformally invariant for arbitrary $n$ in Ref.~\cite{EdwardWangBSc}.

\section{Hints for dual conformal symmetry}
\label{sec:DualConformalSymmetry}

The numerators of the summands in our formula~\eqref{eq:Ckmn} have an intriguing reminiscence of the NMHV super-amplitudes in $\mathcal{N}=4$ super Yang-Mills. The latter enjoy an additional symmetry, dual (super)conformal symmetry~\cite{Drummond:2008vq}. It is therefore natural to ask whether the one-loop all-plus amplitudes have traces of this symmetry as well.

Dual conformal symmetry is a dynamic symmetry, namely it is not present in the Lagrangian, and emerges only at the level of the scattering amplitudes. Its presence can be most naturally revealed by viewing the scattering amplitudes as functions of the dual coordinates $x_i$, related to the momenta by $p_i = x_i-x_{i+1}$ with $x_{n+1} \equiv x_1$. Then, dual conformal symmetry is the ordinary conformal symmetry in dual space.

We find that the one-loop four-gluon all-plus amplitude is indeed a dual conformal invariant, namely it is annihilated by the infinitesimal generator of dual special conformal transformations~\cite{Drummond:2008vq, Drummond:2009fd},
\begin{align}
\label{eq:dual_conformal_generator}
K^{\alpha \dot{\alpha}} = \sum_{i=1}^n \left[x_i^{\dot{\alpha} \beta} \lambda_i^{\alpha} \frac{\partial}{\partial \lambda_i^{\beta}} + x_{i+1}^{\dot{\beta} \alpha} \tilde{\lambda}^{\dot{\alpha}}_i  \frac{\partial}{\partial \tilde{\lambda}^{\dot{\beta}}_i} \right] \, .
\end{align}
At higher multiplicity, the full all-plus amplitude does not retain any sign of dual conformal symmetry as a whole. Quite remarkably, however, the individual summands $C_{kmn}$ are actually covariant under dual special conformal transformations projected along a specific direction.

The easiest way to see that the one-loop four-gluon all-plus amplitude is dual conformally invariant is to look at the infinitesimal variation generated by $K^{\alpha \dot{\alpha}}$~\eqref{eq:dual_conformal_generator} projected along some direction $b$,
\begin{align}
\delta_b = b_{\dot{\alpha} \alpha} K^{\alpha \dot{\alpha}} \, .
\end{align}
For convenience of the reader we list the infinitesimal transformations of the spinors and of the dual variables:
\begin{align} 
\label{eq:DCT1} & \delta_b |i\rangle = x_i b |i\rangle \, , \\
\label{eq:DCT2} & \delta_b |i] = x_{i+1} b|i] \, ,\\
\label{eq:DCT3} & \delta_b x_{ij} = x_i b x_{ij} + x_{ij} b x_{j} \, .
\end{align}
A function $f$ transforms covariantly with weight $w$ under dual conformal transformations if $\delta_b f = w \, f$. From Eqs.~\eqref{eq:DCT1},~\eqref{eq:DCT2} and~\eqref{eq:DCT3} it is straightforward to see that spinor contractions with adjacent indices transform covariantly under dual conformal symmetry,
\begin{align}
& \delta_b \vev{i\, i+1} = 2 (b \cdot x_i) \vev{i \, i+1} \, , \\
& \delta_b [i\, i+1] = 2 ( b \cdot x_{i+2} ) [i \, i+1] \, .
\end{align}
Using these equations it is easy to verify that the one-loop four-gluon all-plus amplitude given by Eq.~\eqref{eq:A4literature} or~\eqref{A4explicit} enjoys dual conformal symmetry. In fact, it is a Yangian invariant~\cite{Drummond:2009fd}.

At $n=5$ several spinor brackets have non-adjacent indices, therefore the dual conformal properties are not obvious at first sight.
Looking at the general formula~\eqref{eq:our_formula}, we see that $\lambda_{1}$ breaks the dual conformal covariance.
For instance, $\langle 1 | x_{1k} x_{km}|m\rangle$ is dual conformally covariant, but projecting by $|1\rangle$ rather than $|m\rangle$ breaks the symmetry,
\begin{align}
\label{eq:exampleDDCI}
\delta_b \langle 1 | x_{1k} x_{km}| 1\rangle = 2 b\cdot (x_1 + x_k + x_m)  \langle 1 | x_{1k} x_{km}| 1\rangle - \langle 1 | x_{1k} x_{km} x_{m1} b |1\rangle \, .
\end{align}
We recognise on the left-hand side of Eq.~\eqref{eq:exampleDDCI} the factor in the numerator of the summands $C_{kmn}$~\eqref{eq:Ckmn}.
The same conclusion holds for the full $C_{kmn}$. 

The fact that $\lambda_1$ alone breaks the symmetry, however, suggests to project the dual conformal variation along a specific direction, in our case $p_1$\footnote{In fact, it suffices to project by $\lambda_1$. Any parameter $b_{\alpha \dot{\alpha}} \propto \lambda_{1 \alpha} \tilde{\lambda}_{j \dot{\alpha}} \, \forall j$ would eliminate the non-covariant terms. We prefer to project by $b \propto p_1$ for simplicity.}. The existence of such a sub-group of dual conformal symmetry, called directional dual conformal symmetry, has been recently unveiled and used in the computation of certain non-planar loop integrals~\cite{Bern:2017gdk,Bern:2018oao,Chicherin:2018wes} (see also Ref.~\cite{Ben-Israel:2018ckc}).
The projection by $p_1$ in fact removes all the terms which would break the dual conformal covariance, as can be clearly seen by taking $b \propto p_1$ in Eq.~\eqref{eq:exampleDDCI}. This implies that the numerator of $C_{kmn}$~\eqref{eq:Ckmn} transforms in a covariant way under dual conformal variations along the direction \mbox{$b = \eps \, p_1$}, for some infinitesimal parameter $\eps$.
Remarkably, the same holds for the denominator, so that each summand $C_{kmn}$ in our formula~\eqref{eq:our_formula} is directionally dual conformally covariant,
\begin{align}
\label{eq:CovarianceCkmn}
\delta_{\eps p_1} C_{kmn} = 2 \eps \, p_{1} \cdot \left(2 x_{1} + x_{k} + x_{m} - \sum_{i=1}^n x_i \right) \, C_{kmn} \,.
\end{align}

It is clear, however, that each term in our formula~\eqref{eq:our_formula} transforms with different dual conformal weights, so that the entire formula does not have a symmetry that can be clearly stated. If the whole formula had been covariant along some direction, then --~because of the cyclic symmetry of the all-plus planar amplitude~-- it would have been covariant along any direction. This is what happens in the $n=4$ case: the sum contains only one summand, which indeed is exactly dual conformal.
At higher multiplicity, the presence of a preferred direction, $p_1$, is a mere artefact of the representation of the summands we chose, but has no meaning for the complete amplitude. 
Nevertheless, it is remarkable that the one-loop all-plus amplitude can be written in such a form that each term, separately, exhibits dual conformal covariance properties.

\begin{comment}
This follows from the properties of BCFW recursion. In Ref.~\cite{Brandhuber:2008pf}, the dual (super)conformal symmetry of tree-level super-amplitudes in $\mathcal{N}=4$ super Yang-Mills was proven by recursion using a BCFW shift of two adjacent momenta. The super-momentum conservation $\delta$ function there plays an important role, by balancing the weights of the separate BCFW diagrams so that the super-amplitude as a whole maintains its covariance in the recursion. By restricting to the bosonic part, namely to tree-level gluon amplitudes, the weights of the separate BCFW diagrams remain unbalanced, and the dual conformal symmetry of the four-gluon amplitude is lost at higher multiplicity.

We see similar expressions explains the occurrence of similar hints of dual conformal symmetry in various other contexts as well (e.g. see Eq.~(2.8) of Ref.~\cite{Stieberger:2016lng}, and Ref.~\cite{Drummond:2009ge}).
\end{comment}

\section{Analytic structure of the amplitude}
\label{sec:FactorizationSpurious}

There is a tension between manifest conformal symmetry and manifest analytic properties. As discussed in Section~\ref{sec:PresentationResult}, our formula~\eqref{eq:our_formula} is manifestly conformally invariant, but its analytic properties are obscured. 
In this section we verify that it has the correct leading behaviour in the soft and collinear limits, and that the spurious poles where non-adjacent momenta become collinear have vanishing residues.

Note that the collinear behaviour and the absence of spurious poles fix entirely the one-loop all-plus amplitudes for $n>5$~\cite{Bern:1993sx, Bern:1993qk}. The fact that our formula~\eqref{eq:our_formula} matches the known $n=4,5$ results and has the correct analytic properties is a proof of its validity.

\subsection{Soft limit}
\label{sec:Soft}

Let us first consider the soft limit, namely the limit in which all components of a gluon momentum, say $p_j$, go to zero. The leading divergence of the one-loop all-plus planar amplitude factorises into a universal tree-level soft factor and the one-loop amplitude with gluon $j$ removed~\cite{Weinberg:1964, Low:1958},
\begin{equation}\label{eq:soft_behaviour}
    A^{(1,1)}_n(1^+,\ldots,n^+) \, \overset{p_j\rightarrow 0}{\sim} \, \text{Soft}^{(0)}\left(j-1,j^+,j+1 \right) A^{(1,1)}_{n-1}(1^+,\ldots,\hat{j}^+,\ldots,n^+) \, .
\end{equation}
where
\begin{align}
\text{Soft}^{(0)}\left(i,j^+,k \right) = \frac{\vev{ik}}{\vev{ij} \vev{jk}} \, .
\end{align}
Note that, for a generic one-loop amplitude, Eq.~\eqref{eq:soft_behaviour} would contain also a term with the one-loop soft factor and the tree-level amplitude, but the latter vanishes in the all-plus helicity case.

Let us now investigate the leading soft behaviour of our formula~\eqref{eq:our_formula}. For this purpose, it is convenient to start from Eq.~\eqref{eq:CkmnAlternative} for the summands $C_{kmn}$. The terms with \mbox{$k=j$}, \mbox{$m=j+1$} in the sum vanish as $p_j \to 0$. The others give
\begin{equation}\label{eq:soft}
    \begin{aligned}
       A_{n}^{(1,1)}&(1^+,\ldots ,n^+) \underset{p_j\to 0}{\sim} \frac{\vev{j-1 \, j+1}}{\vev{j-1 \, j}\vev{j \, j+1}} \frac{-M}{\PT(1,\ldots,\hat{j},\ldots,n)} \\
          & \bigg\lbrace \bigg( \sum_{k=3}^{j-1} \sum_{m=k+1}^{j-1} + \sum_{k=3}^{j-1} \sum_{m=j+2}^{n} + \sum_{k=j+2}^{n-1} \sum_{m=k+1}^{n} \bigg) \frac{\langle 1 \vert x_{1 k } x_{k m} \vert 1 \rangle ^2 \ab{k-1}{k} \ab{m-1}{m}}{\ab{1}{k-1} \ab{1}{k} \ab{1}{m-1} \ab{1}{m}}  \\
         & + \frac{\vev{j-1\, j+1}}{\vev{1 \, j-1}\vev{1 \, j+1}} \bigg(\sum_{k=3}^{j-1} \frac{\langle 1 \vert x_{1k} x_{k, j+1} \vert 1 \rangle ^2 \ab{k-1}{k}}{\ab{1}{k-1} \ab{1}{k}} \\
        & +  \sum_{m=j+2}^{n} \frac{\langle 1 \vert x_{1, j+1} x_{j+1, m} \vert 1 \rangle ^2 \ab{m-1}{m}}{\ab{1}{m-1} \ab{1}{m}} \bigg) \bigg\rbrace \, .
    \end{aligned}
\end{equation}
We see that the single sums in Eq.~\eqref{eq:soft} extend the second double sum to $m=j+1$, and the third to  $k=j+1$, respectively.
The resulting expression is therefore that of the one-loop $(n-1)$-gluon all-plus amplitude $A^{(1,1)}_{n-1}(1^+, \ldots, \hat{j}^+, \ldots, n^+ )$, in perfect agreement with the expected leading soft behaviour~\eqref{eq:soft_behaviour}.

\subsection{Collinear limit}
We now check that our expression~\eqref{eq:our_formula} exhibits the correct behaviour as two adjacent momenta, say $p_j$ and $p_{j+1}$, become collinear, i.e.
\begin{equation}
\label{eq:collinear_limit}
	\begin{aligned}
		& p_j \to z \, P\, ,  \\
		& p_{j+1} \to (1-z) \, P \, ,
	\end{aligned}
\end{equation}
for some factor $z$, with $P = p_j + p_{j+1}$.

At one loop, the leading behaviour of the planar $n$-gluon all-plus amplitude in the collinear limit~\eqref{eq:collinear_limit} has the form~\cite{Mangano:1991, PhysRevLett.56.2459, BERENDS1988759, MANGANO1988673,Bern:1994zx}
\begin{equation}\label{eq:collinear_factorization}
	\begin{aligned}
		A^{(1,1)}_n(\ldots, j^+, (j+1)^+, \ldots) \overset{\ \ j \parallel j+1}{\sim} \ \text{Split}^{(0)}_{-}\left(z; j^{+}, (j+1)^{+}\right) \, A^{(1,1)}_{n-1}\left(\ldots, P^{+},\ldots\right)  \, ,
	\end{aligned}
\end{equation}
where $\text{Split}^{(0)}_-$ is the tree-level $g^-\to g^+g^+$ splitting function,
\begin{equation}
\text{Split}^{(0)}_{-} \left(z; j^{+}, (j+1)^+\right) = \frac{1}{\sqrt{z(1-z)} \ab{j}{j+1}} \,.
\end{equation}
Just like in the soft limit case, the behaviour of the one-loop all-plus amplitude is particularly simple, because the all-plus and single-minus amplitudes vanish at tree-level. In general, there would be a contribution from the one-loop splitting functions and the tree-level amplitudes as well.

In order to take the collinear limit of our expression (\ref{eq:our_formula}), we follow the procedure described in Ref.~\cite{Stieberger:2015}. We introduce two auxiliary momenta, $P = \lambda_P \tilde{\lambda}_P$ and $r = \lambda_r \tilde{\lambda}_r$, and two parameters, $\theta$ and $\epsilon$, to parametrise the spinor helicity variables as
\begin{equation}\label{collinear_limit}
	\begin{aligned}
		& \spn{j} = \spn{P} \text{cos}\theta - \epsilon \spn{r} \text{sin}\theta \,, \qquad & \asp{j} = \asp{P} \text{cos}\theta - \epsilon \asp{r} \text{sin}\theta \, ,\\
		& \spn{j+1} = \spn{P} \text{sin}\theta + \epsilon \spn{r} \text{cos}\theta \,, \qquad & \asp{j+1} = \asp{P} \text{sin}\theta + \epsilon \asp{r} \text{cos}\theta \, ,
	\end{aligned}	
\end{equation}	
such that
\begin{equation}	
	p_j + p_{j+1} = \spn{P} \asp{P} + \epsilon ^2 \spn{r} \asp{r} \, .
\end{equation}
In the limit $\epsilon \rightarrow 0$ we recover Eq.~\eqref{eq:collinear_limit}, with $\text{cos}\theta = \sqrt{z}$ and $\text{sin}\theta = \sqrt{1-z}$.

The collinear analysis of our formula~\eqref{eq:our_formula} is then very similar to the soft one, and reveals perfect agreement with the expected behaviour~\eqref{eq:collinear_factorization}.

\subsection{Absence of spurious poles}

Finally, we know that the planar amplitude does not contain poles where two non-adjacent momenta become collinear. Let us show that Eq.~(\ref{eq:our_formula}) is free of such poles. Looking at the expression of the summands given by Eq.~\eqref{eq:CkmnAlternative}, it is clear that the only spurious poles are of the type $1/\vev{1j}$. The terms of the sum in Eq.~(\ref{eq:our_formula}) that appear to contain this pole are the ones with $k,k-1 = j$ or $m, m-1 = j$. The term $C_{j,j+1,n}$ does not actually contain any $1/\ab{1}{j}$ pole, as becomes apparent on rewriting
\begin{align*}
	C_{j,j+1,n}  =  - \frac{1}{\PT(1,2,\ldots,n)} \frac{ \langle 1 \vert x_{1 j} \vert j ] ^2 \vev{j-1 \, j} \vev{j \, j+1}}{\ab{1}{j-1} \ab{1}{j+1}} \, ,
\end{align*}
starting from Eq.~\eqref{eq:CkmnAlternative}.
The remaining terms contributing to the pole are then
\begin{equation}\label{spur}
	\begin{aligned}
		\frac{M \, B}{\PT(1,\ldots,n)} \bigg( \sum_{m=j+2}^n \frac{\langle 1 \vert x_{1,j+1} x_{j+1,m} \vert 1 \rangle ^2 \ab{m-1}{m}}{\ab{1}{m-1} \ab{1}{m}} + \sum_{k=3}^{j-1} \frac{\langle 1 \vert x_{1,k} x_{k,j+1} \vert 1 \rangle ^2 \ab{k-1}{k}}{\ab{1}{k-1} \ab{1}{k}} \bigg) \, ,
	\end{aligned}
\end{equation}
where
\begin{align}
B = \frac{\vev{j-1 \, j}}{\vev{1 \, j-1} \vev{1 \, j}} + \frac{\vev{j \, j+1}}{\vev{1 \, j}\vev{1 \, j+1}} \,.
\end{align}
Through Schouten identities it is however easy to see that
\begin{align}
B = \frac{\vev{j-1 , j+1}}{\vev{1\, j-1}\vev{1 \, j+1}} \, ,
\end{align}
so that no $1/\ab{1}{j}$ pole is left.

\section{Proof of the new formula via BCFW recursion}
\label{sec:BCFW}

On-shell recursive techniques have proven very successful for the computation of scattering amplitudes. In this section we prove our all-$n$ formula~\eqref{eq:our_formula} for the one-loop all-plus amplitudes through the Britto-Cachazo-Feng-Witten (BCFW) recursion~\cite{Britto:2005}. We start with a briew review of this technique.

If an $n$-particle scattering amplitude $A_n$ is finite and rational, its form can be reconstructed from the knowledge of its poles and residues. The idea at the basis of the BCFW recursion is to perform such analysis by shifting by a complex parameter $z$ the external momenta, and studying the amplitude evaluated at the deformed kinematics $A_n(z)$ as a function of $z$. 

The momenta, say $p_a$ and $p_b$, are shifted in such a way that on-shellness and momentum conservation are preserved. This can be achieved for instance by shifting the corresponding spinors as
\begin{equation}\label{eq:generic_shift}
    \asp{a} \rightarrow \asp{a} - z \asp{b} \,, \qquad \spn{b} \rightarrow \spn{b} + z \spn{a} \,.
\end{equation}
Then, if the shifted amplitude $A_n(z)$ is a finite rational function with poles at points $z_i$, we can apply Cauchy's theorem to $A_n(z)/z$ with a contour encircling infinity. As a result, the original unshifted amplitude $A_n = A_n \left(z=0\right)$ is rewritten in terms of the residues of the shifted one,
\begin{equation}\label{eq:cauchy}
    A_n = C^{\infty}_n - \sum_{\text{poles } z_i} \underset{z = z_i}{\text{Res}}\left( \frac{A_n(z)}{z} \right) \,,
\end{equation}
where the sum runs over the finite-distance poles $z_i$, i.e. $|z_i|<\infty$, and $C^{\infty}_n$ is the residue at $z=\infty$, i.e.
\begin{equation}
    C^{\infty}_n = \lim_{z\rightarrow \infty} A_n(z) \,.
\end{equation}

Tree-level amplitudes are rational functions of the spinor products, and hence of $z$. Furthermore, they have only simple poles, corresponding to multi-particle and collinear poles. On such poles a tree-level amplitude $A^{(0)}_n$ factorises into the product of lower-multiplicity tree amplitudes, so that Eq.~\eqref{eq:cauchy} becomes
\begin{equation}
\label{eq:BCFWrecursion}
    A^{(0)}_n = C^{\infty}_n  + \sum_{i,\sigma} A^{(0)\, \sigma}_{n_i}(z=z_i)\frac{i}{P_i^2} A^{(0) \, -\sigma}_{n+2-n_i}(z=z_i) \,,
\end{equation}
where $\sigma = \pm 1$ is the helicity of the intermediate state with momentum $P_i$, and $i$ labels all possible factorisations such that the shifted legs $a$ and $b$ appear on opposite sides of the pole.
Finally, at tree level it is always possible to find a choice for the shifted momenta~\eqref{eq:generic_shift} such that there is no pole at infinity. In such case, Eq.~\eqref{eq:BCFWrecursion} gives a completely algorithmic recursion relation for the amplitude.

At loop level the situation is more complicated. First of all, loop amplitudes are in general not rational. The all-plus and single-minus amplitudes however vanish at tree-level, so that they are finite and rational at one loop. 

The second complication is due to the potential presence of double poles, introduced by the one-loop all-plus and all-minus splitting functions regulating the collinear singularities~\cite{Bern:1994zx}. The latter are however removed by vanishing tree-level amplitudes, so that the one-loop all-plus amplitude has only simple poles. Note however that the one-loop single-minus amplitude, instead, does contain double poles. 

Thanks to these simplifications, the BCFW recursion relation for the one-loop all-plus amplitudes is identical to the one for tree-level amplitudes given by Eq.~\eqref{eq:BCFWrecursion}, with an additional sum over the two ways of assigning the loop to the pair of lower-multiplicity amplitudes,
\begin{equation}
\label{eq:BCFWrecursionOneLoop}
    A^{(1)}_n = C^{\infty}_n  + \sum_{\ell=0,1} \sum_{i,\sigma} A^{(1-\ell)\, \sigma}_{n_i}(z=z_i)\frac{i}{P_i^2} A^{(\ell) \, -\sigma}_{n+2-n_i}(z=z_i) \,.
\end{equation}

Finally, there is the issue of the residue at infinity, which produces the surface term~$C^{\infty}_n$ in the recursions~\eqref{eq:BCFWrecursion} and~\eqref{eq:BCFWrecursionOneLoop}. As can be understood from the known result~\eqref{allplus_all_n}, there is no shift of the form~\eqref{eq:generic_shift} for which the amplitude would vanish as $z\to \infty$. In Ref.~\cite{Bern:2005hs} the authors propose a more refined shift of three spinors which removes the pole at infinity. However, that shift breaks the manifest conformal invariance of the separate terms in the recursion.

The ordinary BCFW shift of the form~\eqref{eq:generic_shift}, on the other hand, is conformally invariant: any term constructed via BCFW recursion with shift~\eqref{eq:generic_shift} from conformally invariant lower-multiplicity amplitudes is in its turn conformally invariant. 
This remarkable property becomes transparent in twistor space~\cite{Mason:2009sa}. 

Since our goal is to expose the conformal invariance of the one-loop all-plus amplitude, we prefer to tame the pole at infinity by combining two complementary BCFW shifts~\cite{Berger:2006ci,Dunbar:2010wu}.
The basic idea is that the residue at infinity for one shift may be given by finite-distance residues of a second, auxiliary shift. By combining two suitably chosen BCFW shifts, therefore, it may be possible to reconstruct the whole amplitude, including the surface term.

We now present a proof by induction of our formula~\eqref{eq:our_formula}. The proof is based on the recursion relation~\eqref{eq:BCFWrecursionOneLoop} coming from the shift
\begin{align}\label{shift21}
	\lambda_2 \longrightarrow \hat{\lambda}_2 = \lambda_2 + z \lambda_1 \,,\qquad \quad \tilde\lambda_{1} \longrightarrow \hat{\tilde{\lambda}}_{1} = \tilde\lambda_{1} - z \tilde\lambda_{2} \, ,
\end{align}
which we denote by $\langle2,1]_z$. In section~\ref{sec:TermInfinity} we show that the amplitude has a non-vanishing residue at $z \to \infty$ under the deformation $\langle 2,1]_z$, so that the recursion relation contains a surface term. We find that this surface term is contained in the finite-distance residues of a second, auxiliary BCFW shift,
\begin{align}\label{shiftn1}
	\lambda_{n-1} \longrightarrow \hat{\lambda}_{n-1} = \lambda_{n-1} + w \lambda_1 \,,\qquad \quad \tilde\lambda_{1} \longrightarrow  \hat{\tilde{\lambda}}_{1} = \tilde\lambda_{1} - w \tilde\lambda_{n-1} \,,
\end{align}
labeled by $\langle n-1, 1]_w$. We then use the latter shift to determine by recursion the surface term of the original recursion based on $\langle2,1]_z$. Finally, we solve the recursion for the complete one-loop all-plus amplitude in section~\ref{sec:Induction}.

\subsection{Term at infinity}
\label{sec:TermInfinity}
Let us consider the behaviour at $z\to \infty$ of our formula~\eqref{eq:our_formula} under the deformation $\langle 2,1]_z$~\eqref{shift21}.
The numerator in Eq.~(\ref{eq:our_formula}) is independent of $z$ thanks to the projection by $\lambda_{1}$.
Analysing the denominator, we see that one type of term leads to a non-vanishing behaviour at infinity. The ensuing residue at infinity is 
\begin{equation}\label{eq:C3m}
	\begin{aligned}
		C^{ \infty }_{n} = & \lim_{z \to \infty} A^{(1,1)}_{n}(z) \, \\
		=& M \sum_{m=4}^{n} \frac{ [2| x_{3n} |1\rangle^2 }{\PT(1, 3, 4, \ldots, m-1 ) \, \PT(1, m, m+1, \ldots , n )} \, \\
		=& \, \frac{M}{\PT(1,3,\ldots,n)} \sum_{m=4}^{n} \frac{ [2 \vert x_{3n} \vert 1 \rangle ^2 \ab{m-1}{m}}{\ab{m-1}{1} \ab{1}{m}} \,. 		
	\end{aligned}
\end{equation}
Note that, after some manipulation and the use of telescoping series, this can be shown to match the residue at infinity obtained from the original formula~\eqref{allplus_all_n}\footnote{Note that one could compute the surface term from the known all-$n$ formula~\eqref{allplus_all_n} and plug it into the recursion~\eqref{eq:BCFWrecursionOneLoop} (e.g. see~\cite{He:2014bga}). We prefer to show a derivation from first principles, i.e. that does not assume Eq.~\eqref{allplus_all_n}.}. In this section we show that the surface term $C^{\infty}_n$~\eqref{eq:C3m} can be obtained via the auxiliary BCFW recursion $\langle n-1, 1]_w$~\eqref{shiftn1}, without assuming the knowledge of the full amplitude.

The surface term given by Eq.~\eqref{eq:C3m} vanishes at infinity under certain other shifts, in particular by deforming momenta $p_1$ and $p_{n-1}$. It is thus natural to expect that we can find $C^{\infty}_{n}$  from $C^{\infty}_{n-1}$ using the BCFW shift $\langle n-1, 1]_w$~\eqref{shiftn1}.

In general there is some overlap between the finite-distance residues arising from two different shifts, and --~provided that the shifts complement each other~-- the full solution can be reconstructed by careful comparison. Doing this with the two shifts $\langle2,1]_z$~\eqref{shift21} and $\langle n-1, 1]_w$~\eqref{shiftn1} for $n=5, 6, 7$, a clear pattern emerges: the $n$-point surface term $C^{\infty}_n$ is given by the residues at the finite-distance poles of the  $(n-1)$-surface term $C^{\infty}_{n-1}$ under the shift $\langle n-1, 1]_w$~\eqref{shiftn1}.

\begin{figure}[t]
	\begin{subfigure}{0.5\textwidth}
  		\begin{tikzpicture}
    		\draw [line width=0.15mm] (3,0) circle (0.8cm);
    		\draw [line width=0.15mm] (5.8,0) circle (0.4cm);
    		\draw [line width=0.15mm] (3,0) circle (0.5cm);
    		\draw [directed, line width=0.15mm] (5.4,0) -- (3.8,0);
    		\draw [directed, line width=0.15mm] (7,0.95) -- (6.1,0.28);
    		\draw [directed, line width=0.15mm] (7,-0.95) -- (6.1,-0.28);
    		\draw [directed, line width=0.15mm] (2.4,1.85) -- (2.8,0.78);
    		\draw [directed, line width=0.15mm] (1.55,-1.4) -- (2.4,-0.55);
    		\draw [directed, line width=0.15mm] (2.4,-1.85) -- (2.8,-0.78);
			\node[] at (7.4,1.2)  {$(\widehat{n-1})^+$};
			\node[] at (7.4,-1.1)  {$n^+$};
			\node[] at (2.4,2.1)  {$(n-2)^+$};
			\node[] at (1.4,-1.6)  {$2^+$};
			\node[] at (2.4,-2.1)  {$\hat{1}^+$};
			\node[] at (4.6,-0.3)  {$\hat{P}_{n-1,n}$};
			\node[] at (4,0.2)  {$+$};
			\node[] at (5.2,0.2)  {$-$};
			\node[] at (5.8,0)  {$\overline{\text{\tiny MHV}}$};
			\node[] at (1.7,0.3)  {$\cdot$};
			\node[] at (1.8,-0.5)  {$\cdot$};
			\node[] at (2.1,1)  {$\cdot$};
			\node[] at (3,0) {$C^{\infty}_{n-1}$};
		\end{tikzpicture}
		\caption{}
		\label{fig:BCFW_1n_a}
	\end{subfigure}
	\begin{subfigure}{0.5\textwidth}
  		\begin{tikzpicture}
    		\draw [line width=0.15mm] (3,0) circle (0.8cm);
    		\draw [line width=0.15mm] (5.8,0) circle (0.4cm);
    		\draw [line width=0.15mm] (3,0) circle (0.5cm);
    		\draw [directed, line width=0.15mm] (5.4,0) -- (3.8,0);
    		\draw [directed, line width=0.15mm] (7,0.95) -- (6.1,0.28);
    		\draw [directed, line width=0.15mm] (7,-0.95) -- (6.1,-0.28);
    		\draw [directed, line width=0.15mm] (2.4,1.85) -- (2.8,0.78);
    		\draw [directed, line width=0.15mm] (1.6,-1.4) -- (2.47,-0.63);
    		\draw [directed, line width=0.15mm] (2.4,-1.85) -- (2.8,-0.78);
    		\draw [directed, line width=0.15mm] (1.2,-0.75) -- (2.23,-0.25);
			\node[] at (7.4,1.2)  {$(n-2)^+$};
			\node[] at (7.4,-1.2)  {$(\widehat{n-1})^+$};
			\node[] at (2.4,2.1)  {$(n-3)^+$};
			\node[] at (1.35,-1.6)  {$\hat{1}^+$};
			\node[] at (2.4,-2.1)  {$n^+$};
			\node[] at (4.6,-0.3)  {$\hat{P}_{n-2,n-1}$};
			\node[] at (1,-0.9)  {$2^+$};
			\node[] at (4,0.2)  {$+$};
			\node[] at (5.2,0.2)  {$-$};
			\node[] at (5.8,0)  {$\overline{\text{\tiny MHV}}$};
			\node[] at (1.85,-0.05)  {$\cdot$};
			\node[] at (1.8,0.5)  {$\cdot$};
			\node[] at (2.15,0.95)  {$\cdot$};
			\node[] at (3,0) {$C^{\infty}_{n-1}$};
		\end{tikzpicture}
		\caption{}
		\label{fig:BCFW_1n_b}
	\end{subfigure}
	\caption{Diagrams contributing to the recursion~\eqref{eq:RecursionCn} for the surface term $C_n^{\infty}$ based on the BCFW shift $\langle n-1, 1]_w$~\eqref{shiftn1}.}
	\label{fig:BCFW_1n}
\end{figure}
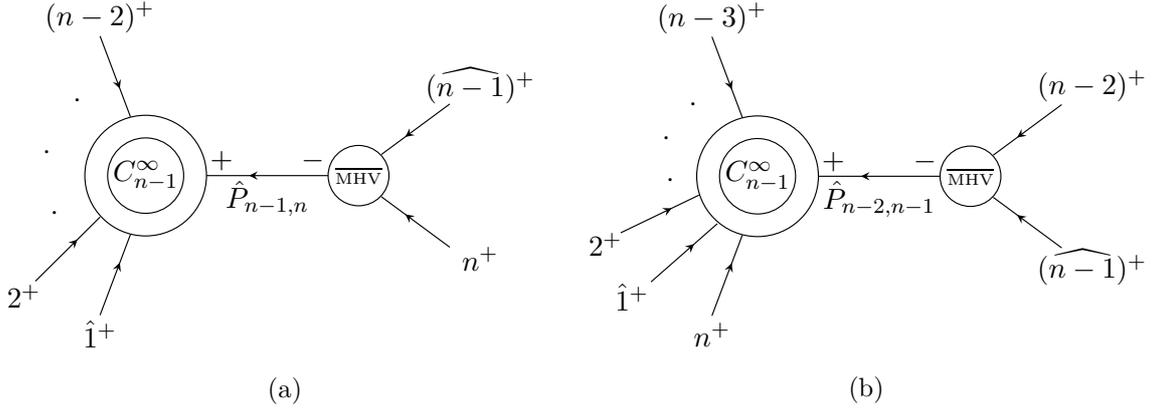

This translates into a homogeneous recursion relation for the surface term $C^{\infty}_n$, based on the BCFW shift $\langle n-1, 1]_w$~\eqref{shiftn1}, which is completely decoupled from the rest of the amplitude. This recursion receives contributions from the two diagrams shown in Fig.~\ref{fig:BCFW_1n}, and explicitly reads
\begin{equation}
\label{eq:RecursionCn}
\begin{aligned}
C_n^{\infty} = \ & C^{\infty}_{n-1}(\hat{1},2,\ldots, n-2,\hat{P}_{n-1,n}) \frac{i}{P^2_{n-1,n}} A_{\overline{\text{MHV}}}(\widehat{n-1},n,-\hat{P}_{n-1,n}) \\
& + C^{\infty}_{n-1}(\hat{1},2,\ldots, n-3,\hat{P}_{n-2,n-1},n) \frac{i}{P^2_{n-2,n-1}} A_{\overline{\text{MHV}}}(n-2,\widehat{n-1},-\hat{P}_{n-2,n-1}) \,, 
\end{aligned}
\end{equation}
where the hats denote deformation according to Eq.~\eqref{shiftn1}, $P_{ij} = p_i + p_j$, and $A_{\overline{\text{MHV}}}$ is the three-point anti-MHV amplitude,
\begin{align}
\label{eq:A3antiMHV}
A_{\overline{\text{MHV}}}(a,b,c) = i \frac{[ab]^3}{[bc] [ca]} \, .
\end{align}
It is understood that the two terms on the right-hand side of Eq.~\eqref{eq:RecursionCn} are evaluated at the values of the deformation parameter $w$ for which $\hat{P}^2_{n-1,n} = 0$ and $\hat{P}^2_{n-2,n-1} = 0$, respectively.

We assume, by induction hypothesis, that the $(n-1)$-point surface term $C_{n-1}^{\infty}$ is given by Eq.~\eqref{eq:C3m}. Then, the contribution from the diagram in Fig.~\ref{fig:BCFW_1n_a}, in the first line of Eq.~\eqref{eq:RecursionCn}, can be written as 
\begin{equation}\label{contrib1}
	\begin{aligned}
		\frac{-M}{\PT(1,3,\ldots,n)} \bigg\lbrace \frac{\vev{n-2,n-1} \ab{n}{1}}{\vev{n-2,n} \vev{1,n-1}} \bigg( \sum_{m=4}^{n-2} \frac{ [ 2 \vert x_{3 m} \vert 1 \rangle ^2 \vev{m-1,m}}{\vev{m-1,1} \ab{m}{1}} \bigg) \\ 
		- \frac{ [ 2 \vert x_{3,n-1} \vert 1 \rangle ^2 \vev{n-2,n-1}}{\vev{n-2,1} \vev{n-1,1}} \bigg\rbrace.
	\end{aligned}
\end{equation}
The contribution from the diagram in Fig.~\ref{fig:BCFW_1n_b}, in the second line of Eq.~\eqref{eq:RecursionCn}, is
\begin{equation}\label{contrib2}
	\begin{aligned}
		\frac{-M}{\PT(1,3,\ldots,n)} \bigg\lbrace \frac{\vev{1,n-2} \vev{n,n-1}}{\vev{n-2,n} \vev{1,n-1}} \bigg( \sum_{m=4}^{n-2} \frac{ [ 2 \vert x_{3 m} \vert 1 \rangle ^2 \vev{m-1,m}}{\vev{m-1,1} \ab{m}{1}} \bigg) \\ 
		- \frac{ [ 2 \vert x_{3 n} \vert 1 \rangle ^2 \vev{n-1,n}}{\vev{n-1,1} \ab{n}{1}} \bigg\rbrace.
	\end{aligned}
\end{equation}
Summing Eqs.~(\ref{contrib1}) and~(\ref{contrib2}), and doing some manipulations gives
\begin{align}\label{eq:term_at_infinity}
	\frac{M}{\PT(1,3,\ldots,n)} \sum_{m=4}^{n} \frac{ [ 2 \vert x_{3 m} \vert 1 \rangle ^2 \vev{m-1\,m}}{\vev{m-1\,1} \ab{m}{1}}\, ,
\end{align}
which is exactly $C^{\infty}_n$ as given by Eq.~\eqref{eq:C3m}. 

We have therefore determined that the surface term in the recursion~\eqref{eq:BCFWrecursionOneLoop} is given by Eq.~\eqref{eq:C3m}. We can now move on to solve the recursion for the full one-loop all-plus amplitude.

\subsection{Proof by induction of the new formula}
\label{sec:Induction}
We can now show that our formula~\eqref{eq:our_formula} is the solution of the BCFW recursion relation for the one-loop all-plus amplitude with shift $\langle1, 2]_z$~\eqref{shift21}.

\begin{figure}[t]
	\centering
  	\begin{tikzpicture}
    	\draw [line width=0.15mm] (3,0) circle (0.8cm);
    	\draw [line width=0.15mm] (5.8,0) circle (0.4cm);
    	\draw [line width=0.15mm] (3,0) circle (0.4cm);
    	\draw [directed, line width=0.15mm] (5.4,0) -- (3.8,0);
    	\draw [directed, line width=0.15mm] (7,0.95) -- (6.1,0.28);
    	\draw [directed, line width=0.15mm] (7,-0.95) -- (6.1,-0.28);
    	\draw [directed, line width=0.15mm] (2.4,1.85) -- (2.8,0.78);
    	\draw [directed, line width=0.15mm] (1.55,1.4) -- (2.4,0.55);
    	\draw [directed, line width=0.15mm] (2.4,-1.85) -- (2.8,-0.78);
		\node[] at (7.4,1.2)  {$\hat{2}^+$};
		\node[] at (7.4,-1.1)  {$3^+$};
		\node[] at (2.4,2.1)  {$\hat{1}^+$};
		\node[] at (1.4,1.6)  {$n^+$};
		\node[] at (2.4,-2.1)  {$4^+$};
		\node[] at (4.6,-0.3)  {$\hat{P}$};
		\node[] at (4,0.2)  {$+$};
		\node[] at (5.2,0.2)  {$-$};
		\node[] at (5.8,0)  {$\overline{\text{\tiny MHV}}$};
		\node[] at (1.7,0.2)  {$\cdot$};
		\node[] at (1.8,-0.5)  {$\cdot$};
		\node[] at (2.2,-1.1)  {$\cdot$};
	\end{tikzpicture}
	\caption{Diagram contributing to the BCFW recursion~\eqref{eq:FinalRecursion} for the one-loop all-plus amplitude. The $n$-gluon one-loop all-plus amplitude factorises into the $(n-1)$-gluon one, on the left, and the three-gluon anti-MHV amplitude, on the right.}
	\label{fig:BCFW_12}
\end{figure}
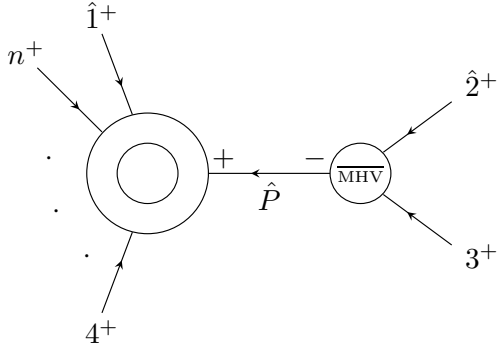

Under the shift $\langle1, 2]_z$~\eqref{shift21}, the recursion relation~\eqref{eq:BCFWrecursionOneLoop} for the one-loop all-plus amplitude receives contribution only from the diagram in Fig.~\ref{fig:BCFW_12},
\begin{align}
\label{eq:FinalRecursion}
	A^{(1,1)}_n(1^+, \ldots, n^+) =   C_n^{\infty}  + A^{(1,1)}_{n-1}(\hat{1},\hat{P},4,\ldots,n-1,n) \frac{i}{P^2} A_{\overline{\text{MHV}}}(\hat{2},3,-\hat{P})\,,
\end{align} 
where the surface term $C_n^{\infty}$ is given by Eq.~\eqref{eq:C3m}, the three-gluon anti-MHV amplitude is defined by Eq.~\eqref{eq:A3antiMHV}, and  $\hat{P} = \hat{p}_2 + p_3$. The deformation parameter $z$ takes the value for which $\hat{P}^2 = 0$.

We assume, by induction hypothesis, that the $(n-1)$-gluon one-loop all-plus amplitude in Eq.~\eqref{eq:FinalRecursion} is given by our formula~\eqref{eq:our_formula}. We spell it out for convenience of the reader,
\begin{equation}
\label{eq:An-1}
\begin{aligned}
	A_{n-1}^{(1,1)}(\hat{1},\hat{P},4,\ldots,n-1,n) = \ & \frac{M}{\PT(1,\hat{P},4,\ldots,n)} \times \\
	& \times \sum_{k=4}^{n-1} \sum_{m=k+1}^{n} \frac{\langle 1 \vert \hat{x}_{1 k} x_{k m} \vert 1 \rangle ^2 \vev{k-1\,k} \vev{m-1\,m}}{\vev{1\,k-1} \ab{1}{k} \vev{1\,m-1} \ab{1}{m}}  \, , \\
\end{aligned}
\end{equation}
where $\hat{x}_{1k} = \hat{p}_1 + \hat{P} +x_{4k}$. Substituting Eq.~\eqref{eq:An-1} into Eq.~\eqref{eq:FinalRecursion} and performing standard manipulations,
\begin{eqnarray}
	\ab{1}{\hat{P}} \sqb{\hat{P}}{3} =  \langle 1 \vert \hat{P} \vert 3 ]  =  \ab{1}{2} \sqb{2}{3} \, , \\
	\sqb{2}{\hat{P}} \ab{\hat{P}}{4} =  [ 2 \vert \hat{P} \vert 4 \rangle  =  \sqb{2}{3} \ab{3}{4} \, ,
\end{eqnarray}
lead to 
\begin{equation}
\label{eq:proofStep}
\begin{aligned}
	A^{(1,1)}_n(1^+, \ldots, n^+) & =  C_n^{\infty}  + \frac{M}{\PT(1,2,\ldots,n)} \sum_{k=4}^{n-1} \sum_{m=k+1}^{n} \frac{\langle 1 \vert x_{1 k} x_{k m} \vert 1 \rangle ^2 \ab{k-1}{k} \ab{m-1}{m}}{\ab{1}{k-1} \ab{1}{k} \ab{1}{m-1} \ab{1}{m}} = \\
	& =   C_n^{\infty}  + M \sum_{k=4}^{n-1} \sum_{m=k+1}^n C_{kmn}  \, , 
\end{aligned}
\end{equation}
where we have recognised in the double sum the $C_{kmn}$ summands as given by Eq.~\eqref{eq:CkmnAlternative}.
Finally, we note that the surface term given by Eq.~\eqref{eq:C3m} can be rewritten as
\begin{equation}
\label{eq:C3mRewriting}
	C_n^{\infty} = \frac{M}{\PT(1,2,\ldots,n)} \sum_{m=4}^{n} \frac{\langle 1 \vert x_{13} x_{3n} \vert 1 \rangle ^2 \ab{2}{3} \vev{m-1\,m}}{\ab{1}{2} \ab{1}{3} \vev{m-1\,1} \ab{1}{m}}  = M \sum_{m=4}^n C_{3mn} \, .
\end{equation}
Substituting Eq.~\eqref{eq:C3mRewriting} into Eq.~\eqref{eq:proofStep} produces our formula~\eqref{eq:our_formula} and completes the proof.

\section{Conclusion and outlook}
\label{sec:Conclusions}

We showed that the all-plus amplitudes in QCD are conformally invariant at one loop. 
The new formula, given by Eq.~\eqref{eq:our_formula}, together with our proof of the conformal invariance of each summand, makes this manifest for arbitrary number of gluons. This constitutes the main result of our paper.
Remarkably, the four-gluon case exhibits also dual conformal symmetry. Together, conformal and dual conformal symmetry make the four-gluon one-loop all-plus amplitude a Yangian invariant. Traces of this propagate also to higher multiplicity: the separate terms of Eq.~\eqref{eq:our_formula} transform in a covariant way under directional dual conformal symmetry.

Our result has consequences at two loops. The rational factors in the polylogarithmic part of the two-loop all-plus amplitudes are known in the planar limit for any number of gluons~\cite{Dunbar:2016cxp}. They are given by on-shell diagrams involving tree-level amplitudes and one-loop all-plus amplitudes as vertices. Therefore, the conformal invariance of the one-loop all-plus amplitudes implies that of these two-loop rational factors. We expect the same to hold for the non-planar contributions as well, as can already be seen in the recently computed full two-loop five-gluon all-plus amplitude~\cite{Badger:2019djh,Dunbar:2019fcq}.

This work is one step towards unraveling the consequences of conformal symmetry for scattering amplitudes.
The latter is in general obscured at quantum level due to ultraviolet and infrared divergences, and due to the on-shell effects revealed by Refs.~\cite{Witten:2003nn,Chicherin:2017bxc,Chicherin:2018ubl,Chicherin:2018rpz}. Nevertheless, we have shown that the all-plus amplitudes are conformally invariant at one loop. 
Interestingly, we find that the one-loop single-minus amplitudes are not conformally invariant, although finite and rational like the all-plus case. Uncovering the underlying conformal properties of the single-minus amplitudes will be an important milestone.

More generally, we would like to understand how conformal symmetry is implemented in the presence of ultraviolet and infrared divergences. 
We consider this avenue to be of great interest for future studies.

\section*{Acknowledgments}
We thank D.~Chicherin, J.~H.~Godwin, D.~Kosower, T.~Peraro, J.~Plefka, E.~Sokatchev and J.~Strong for useful discussions. Preliminary work on conformal invariants for all-plus amplitudes was done in E.~Wang's BSc thesis~\cite{EdwardWangBSc}. B.~P. is supported by the German Academic Exchange Service (DAAD) under the funding programme \textit{Study Scholarships for Graduates of All Disciplines} (No 57381416) and the Elite Network of Bavaria.
This research received funding from the European Research Council (ERC) under the European Union's Horizon 2020 research and innovation programme, {\it Novel structures in scattering amplitudes} (grant agreement No 725110).

\bibliography{BibFile}

\providecommand{\href}[2]{#2}\begingroup\raggedright\begin{thebibliography}{10}

\bibitem{Drummond:2008vq}
J.~M. Drummond, J.~Henn, G.~P. Korchemsky, and E.~Sokatchev, {\it {Dual
  superconformal symmetry of scattering amplitudes in N=4 super-Yang-Mills
  theory}},  {\em Nucl. Phys.} {\bf B828} (2010) 317--374,
  [\href{http://arxiv.org/abs/0807.1095}{{\tt arXiv:0807.1095}}].

\bibitem{Collins:1989gx}
J.~C. Collins, D.~E. Soper, and G.~F. Sterman, {\it {Factorization of Hard
  Processes in QCD}},  {\em Adv. Ser. Direct. High Energy Phys.} {\bf 5} (1989)
  1--91, [\href{http://arxiv.org/abs/hep-ph/0409313}{{\tt hep-ph/0409313}}].

\bibitem{Braun:2003rp}
V.~M. Braun, G.~P. Korchemsky, and D.~Müller, {\it {The Uses of conformal
  symmetry in QCD}},  {\em Prog. Part. Nucl. Phys.} {\bf 51} (2003) 311--398,
  [\href{http://arxiv.org/abs/hep-ph/0306057}{{\tt hep-ph/0306057}}].

\bibitem{Braun:2017cih}
V.~M. Braun, A.~N. Manashov, S.~Moch, and M.~Strohmaier, {\it {Three-loop
  evolution equation for flavor-nonsinglet operators in off-forward
  kinematics}},  {\em JHEP} {\bf 06} (2017) 037,
  [\href{http://arxiv.org/abs/1703.09532}{{\tt arXiv:1703.09532}}].

\bibitem{Bargheer:2009qu}
T.~Bargheer, N.~Beisert, W.~Galleas, F.~Loebbert, and T.~McLoughlin, {\it
  {Exacting N=4 Superconformal Symmetry}},  {\em JHEP} {\bf 11} (2009) 056,
  [\href{http://arxiv.org/abs/0905.3738}{{\tt arXiv:0905.3738}}].

\bibitem{Korchemsky:2009hm}
G.~P. Korchemsky and E.~Sokatchev, {\it {Symmetries and analytic properties of
  scattering amplitudes in N=4 SYM theory}},  {\em Nucl. Phys.} {\bf B832}
  (2010) 1--51, [\href{http://arxiv.org/abs/0906.1737}{{\tt arXiv:0906.1737}}].

\bibitem{Beisert:2010gn}
N.~Beisert, J.~Henn, T.~McLoughlin, and J.~Plefka, {\it {One-Loop
  Superconformal and Yangian Symmetries of Scattering Amplitudes in N=4 Super
  Yang-Mills}},  {\em JHEP} {\bf 04} (2010) 085,
  [\href{http://arxiv.org/abs/1002.1733}{{\tt arXiv:1002.1733}}].

\bibitem{Chicherin:2017bxc}
D.~Chicherin and E.~Sokatchev, {\it {Conformal anomaly of generalized form
  factors and finite loop integrals}},  {\em JHEP} {\bf 04} (2018) 082,
  [\href{http://arxiv.org/abs/1709.03511}{{\tt arXiv:1709.03511}}].

\bibitem{Chicherin:2018ubl}
D.~Chicherin, J.~M. Henn, and E.~Sokatchev, {\it {Scattering Amplitudes from
  Superconformal Ward Identities}},  {\em Phys. Rev. Lett.} {\bf 121} (2018),
  no.~2 021602, [\href{http://arxiv.org/abs/1804.03571}{{\tt
  arXiv:1804.03571}}].

\bibitem{Chicherin:2018rpz}
D.~Chicherin, J.~M. Henn, and E.~Sokatchev, {\it {Amplitudes from anomalous
  superconformal symmetry}},  {\em JHEP} {\bf 01} (2019) 179,
  [\href{http://arxiv.org/abs/1811.02560}{{\tt arXiv:1811.02560}}].

\bibitem{Zoia:2018sin}
S.~Zoia, {\it {Conformal Symmetry and Feynman Integrals}},  {\em PoS} {\bf
  LL2018} (2018) 037, [\href{http://arxiv.org/abs/1807.06020}{{\tt
  arXiv:1807.06020}}].

\bibitem{ArkaniHamed:2012nw}
N.~Arkani-Hamed, J.~L. Bourjaily, F.~Cachazo, A.~B. Goncharov, A.~Postnikov,
  and J.~Trnka, {\em {Grassmannian Geometry of Scattering Amplitudes}}.
\newblock Cambridge University Press, 2016.

\bibitem{Cachazo:2008vp}
F.~Cachazo, {\it {Sharpening The Leading Singularity}},
  \href{http://arxiv.org/abs/0803.1988}{{\tt arXiv:0803.1988}}.

\bibitem{Dunbar:2016cxp}
D.~C. Dunbar, G.~R. Jehu, and W.~B. Perkins, {\it {The two-loop n-point
  all-plus helicity amplitude}},  {\em Phys. Rev.} {\bf D93} (2016), no.~12
  125006, [\href{http://arxiv.org/abs/1604.06631}{{\tt arXiv:1604.06631}}].

\bibitem{Badger:2019djh}
S.~Badger, D.~Chicherin, T.~Gehrmann, G.~Heinrich, J.~M. Henn, T.~Peraro,
  P.~Wasser, Y.~Zhang, and S.~Zoia, {\it {Analytic form of the full two-loop
  five-gluon all-plus helicity amplitude}},  {\em Phys. Rev. Lett.} {\bf 123}
  (2019), no.~7 071601, [\href{http://arxiv.org/abs/1905.03733}{{\tt
  arXiv:1905.03733}}].

\bibitem{Dunbar:2019fcq}
D.~C. Dunbar, J.~H. Godwin, W.~B. Perkins, and J.~M.~W. Strong, {\it {Color
  Dressed Unitarity and Recursion for Yang-Mills Two-Loop All-Plus
  Amplitudes}},  \href{http://arxiv.org/abs/1911.06547}{{\tt
  arXiv:1911.06547}}.

\bibitem{Bern:1991aq}
Z.~Bern and D.~A. Kosower, {\it {The Computation of loop amplitudes in gauge
  theories}},  {\em Nucl. Phys.} {\bf B379} (1992) 451--561.

\bibitem{Bern:1993mq}
Z.~Bern, L.~J. Dixon, and D.~A. Kosower, {\it {One loop corrections to five
  gluon amplitudes}},  {\em Phys. Rev. Lett.} {\bf 70} (1993) 2677--2680,
  [\href{http://arxiv.org/abs/hep-ph/9302280}{{\tt hep-ph/9302280}}].

\bibitem{Bern:1993sx}
Z.~Bern, L.~J. Dixon, and D.~A. Kosower, {\it {New QCD results from string
  theory}},  in {\em {International Conference on Strings 93 Berkeley,
  California, May 24-29, 1993}}, pp.~0190--204, 1993.
\newblock \href{http://arxiv.org/abs/hep-th/9311026}{{\tt hep-th/9311026}}.

\bibitem{Mahlon:1993si}
G.~Mahlon, {\it {Multi - gluon helicity amplitudes involving a quark loop}},
  {\em Phys. Rev.} {\bf D49} (1994) 4438--4453,
  [\href{http://arxiv.org/abs/hep-ph/9312276}{{\tt hep-ph/9312276}}].

\bibitem{Bern:1993qk}
Z.~Bern, G.~Chalmers, L.~J. Dixon, and D.~A. Kosower, {\it {One loop N gluon
  amplitudes with maximal helicity violation via collinear limits}},  {\em
  Phys. Rev. Lett.} {\bf 72} (1994) 2134--2137,
  [\href{http://arxiv.org/abs/hep-ph/9312333}{{\tt hep-ph/9312333}}].

\bibitem{Bern:1990ux}
Z.~Bern and D.~A. Kosower, {\it {Color decomposition of one loop amplitudes in
  gauge theories}},  {\em Nucl. Phys.} {\bf B362} (1991) 389--448.

\bibitem{PhysRevLett.56.2459}
S.~J. Parke and T.~R. Taylor, {\it Amplitude for $n$-gluon scattering},  {\em
  Phys. Rev. Lett.} {\bf 56} (Jun, 1986) 2459--2460.

\bibitem{BERENDS1988759}
F.~Berends and W.~Giele, {\it Recursive calculations for processes with n
  gluons},  {\em Nuclear Physics B} {\bf 306} (1988), no.~4 759 -- 808.

\bibitem{MANGANO1988673}
M.~Mangano and S.~J. Parke, {\it Quark-gluon amplitudes in the dual expansion},
   {\em Nuclear Physics B} {\bf 299} (1988), no.~4 673 -- 692.

\bibitem{Mafra:2012kh}
C.~R. Mafra and O.~Schlotterer, {\it {The Structure of n-Point One-Loop Open
  Superstring Amplitudes}},  {\em JHEP} {\bf 08} (2014) 099,
  [\href{http://arxiv.org/abs/1203.6215}{{\tt arXiv:1203.6215}}].

\bibitem{He:2015wgf}
S.~He, R.~Monteiro, and O.~Schlotterer, {\it {String-inspired BCJ numerators
  for one-loop MHV amplitudes}},  {\em JHEP} {\bf 01} (2016) 171,
  [\href{http://arxiv.org/abs/1507.06288}{{\tt arXiv:1507.06288}}].

\bibitem{Witten:2003nn}
E.~Witten, {\it {Perturbative gauge theory as a string theory in twistor
  space}},  {\em Commun. Math. Phys.} {\bf 252} (2004) 189--258,
  [\href{http://arxiv.org/abs/hep-th/0312171}{{\tt hep-th/0312171}}].

\bibitem{Cachazo:2004by}
F.~Cachazo, P.~Svrcek, and E.~Witten, {\it {Gauge theory amplitudes in twistor
  space and holomorphic anomaly}},  {\em JHEP} {\bf 10} (2004) 077,
  [\href{http://arxiv.org/abs/hep-th/0409245}{{\tt hep-th/0409245}}].

\bibitem{EdwardWangBSc}
E.~Wang, {\it {Conformal Properties of All-Plus Scattering Amplitudes}},  {\em
  BSc. thesis, LMU Munich} (2019). Available at
  \href{https://www.researchgate.net/publication/337338833_Conformal_Properties_of_All-Plus_Scattering_Amplitudes}{www.researchgate.net/publication/337338833\_Conformal\_Properties\_of\_}
  \href{https://www.researchgate.net/publication/337338833_Conformal_Properties_of_All-Plus_Scattering_Amplitudes}{All-Plus\_Scattering\_Amplitudes}.

\bibitem{Drummond:2009fd}
J.~M. Drummond, J.~M. Henn, and J.~Plefka, {\it {Yangian symmetry of scattering
  amplitudes in N=4 super Yang-Mills theory}},  {\em JHEP} {\bf 05} (2009) 046,
  [\href{http://arxiv.org/abs/0902.2987}{{\tt arXiv:0902.2987}}].

\bibitem{Bern:2017gdk}
Z.~Bern, M.~Enciso, H.~Ita, and M.~Zeng, {\it {Dual Conformal Symmetry,
  Integration-by-Parts Reduction, Differential Equations and the Nonplanar
  Sector}},  {\em Phys. Rev.} {\bf D96} (2017), no.~9 096017,
  [\href{http://arxiv.org/abs/1709.06055}{{\tt arXiv:1709.06055}}].

\bibitem{Bern:2018oao}
Z.~Bern, M.~Enciso, C.-H. Shen, and M.~Zeng, {\it {Dual Conformal Structure
  Beyond the Planar Limit}},  {\em Phys. Rev. Lett.} {\bf 121} (2018), no.~12
  121603, [\href{http://arxiv.org/abs/1806.06509}{{\tt arXiv:1806.06509}}].

\bibitem{Chicherin:2018wes}
D.~Chicherin, J.~M. Henn, and E.~Sokatchev, {\it {Implications of nonplanar
  dual conformal symmetry}},  {\em JHEP} {\bf 09} (2018) 012,
  [\href{http://arxiv.org/abs/1807.06321}{{\tt arXiv:1807.06321}}].

\bibitem{Ben-Israel:2018ckc}
R.~Ben-Israel, A.~G. Tumanov, and A.~Sever, {\it {Scattering amplitudes -
  Wilson loops duality for the first non-planar correction}},  {\em JHEP} {\bf
  08} (2018) 122, [\href{http://arxiv.org/abs/1802.09395}{{\tt
  arXiv:1802.09395}}].

\bibitem{Weinberg:1964}
S.~Weinberg, {\it Photons and gravitons in $s$-matrix theory: Derivation of
  charge conservation and equality of gravitational and inertial mass},  {\em
  Phys. Rev.} {\bf 135} (Aug, 1964) B1049--B1056.

\bibitem{Low:1958}
F.~E. Low, {\it Bremsstrahlung of very low-energy quanta in elementary particle
  collisions},  {\em Phys. Rev.} {\bf 110} (May, 1958) 974--977.

\bibitem{Mangano:1991}
M.~L. Mangano and S.~J. Parke, {\it Multiparton amplitudes in gauge theories},
  {\em Phys. Rept.} {\bf 200} (May, 1991) 301--367,
  [\href{http://arxiv.org/abs/hep-th/0509223}{{\tt hep-th/0509223}}].

\bibitem{Bern:1994zx}
Z.~Bern, L.~J. Dixon, D.~C. Dunbar, and D.~A. Kosower, {\it {One loop n point
  gauge theory amplitudes, unitarity and collinear limits}},  {\em Nucl. Phys.}
  {\bf B425} (1994) 217--260, [\href{http://arxiv.org/abs/hep-ph/9403226}{{\tt
  hep-ph/9403226}}].

\bibitem{Stieberger:2015}
S.~Stieberger and T.~R. Taylor, {\it {Subleading Terms in the Collinear Limit
  of Yang-Mills Amplitudes}},  {\em Phys. Lett. B} {\bf 750} (2015) 587--590,
  [\href{http://arxiv.org/abs/hep-ph/1508.01116}{{\tt hep-ph/1508.01116}}].

\bibitem{Britto:2005}
R.~Britto, F.~Cachazo, B.~Feng, and E.~Witten, {\it Direct proof of the
  tree-level scattering amplitude recursion relation in yang-mills theory},
  {\em Phys. Rev. Lett.} {\bf 94} (May, 2005) 181602.

\bibitem{Bern:2005hs}
Z.~Bern, L.~J. Dixon, and D.~A. Kosower, {\it {On-shell recurrence relations
  for one-loop QCD amplitudes}},  {\em Phys. Rev.} {\bf D71} (2005) 105013,
  [\href{http://arxiv.org/abs/hep-th/0501240}{{\tt hep-th/0501240}}].

\bibitem{Mason:2009sa}
L.~J. Mason and D.~Skinner, {\it {Scattering Amplitudes and BCFW Recursion in
  Twistor Space}},  {\em JHEP} {\bf 01} (2010) 064,
  [\href{http://arxiv.org/abs/0903.2083}{{\tt arXiv:0903.2083}}].

\bibitem{Berger:2006ci}
C.~F. Berger, Z.~Bern, L.~J. Dixon, D.~Forde, and D.~A. Kosower, {\it
  {Bootstrapping One-Loop QCD Amplitudes with General Helicities}},  {\em Phys.
  Rev.} {\bf D74} (2006) 036009,
  [\href{http://arxiv.org/abs/hep-ph/0604195}{{\tt hep-ph/0604195}}].

\bibitem{Dunbar:2010wu}
D.~C. Dunbar, J.~H. Ettle, and W.~B. Perkins, {\it {Augmented Recursion For
  One-loop Amplitudes}},  {\em Nucl. Phys. Proc. Suppl.} {\bf 205-206} (2010)
  74--79, [\href{http://arxiv.org/abs/1011.0559}{{\tt arXiv:1011.0559}}].

\bibitem{He:2014bga}
S.~He, Y.-t. Huang, and C.~Wen, {\it {Loop Corrections to Soft Theorems in
  Gauge Theories and Gravity}},  {\em JHEP} {\bf 12} (2014) 115,
  [\href{http://arxiv.org/abs/1405.1410}{{\tt arXiv:1405.1410}}].

\end{thebibliography}\endgroup
\bibliographystyle{JHEP}

\end{document}